\documentclass[aps,pra,superscriptaddress,twocolumn]{revtex4}

\usepackage{graphicx} 
\usepackage{amsmath}
 
\newcommand{\beq}{\begin{equation}}
\newcommand{\enq}{\end{equation}}
\newcommand{\bea}{\begin{eqnarray}}
\newcommand{\ena}{\end{eqnarray}}
\newcommand{\la}{\langle}
\newcommand{\ra}{\rangle}
\newcommand{\kk}{\mathbf{k}}
\newcommand{\QQ}{\mathbf{Q}}

\begin{document}

\title{Interband physics in an ultra-cold Fermi gas in an optical lattice}
\author{J.-P. Martikainen}
\email{jpjmarti@nordita.org}
\affiliation{Nordita, 106 91 Stockholm, Sweden}
\affiliation{Department of Physics, University of Helsinki, 
PO Box 64, 00014 University of Helsinki,  Finland}
\author{E. Lundh}
\affiliation{Department of Physics, Ume\aa\,  University, SE-90187,
Ume\aa, Sweden}
\author{T. Paananen}
\affiliation{Department of Physics, University of Helsinki, 
PO Box 64, 00014 University of Helsinki,  Finland}
\date{\today}

\begin{abstract}
We study a gas of
strongly polarized cold fermions in an optical lattice when
the excited $p$-bands are populated. We derive the relevant Hamiltonian
and discuss the expected phase diagram for both repulsive
and attractive interactions. In the parameter regime 
covered here, 
checkerboard anti-ferromagnetic ordering is found to be possible 
for repulsive interactions while for attractive interactions, transitions
between 
different types of paired phases are 
predicted.
\end{abstract}
%\narrowtext
\maketitle

\section{Introduction}
Experiments with ultra-cold Fermi gases in optical lattices
have opened a way 
to study experimentally interacting fermionic systems in a highly tunable 
environment~\cite{Kohl2005a,Chin2006a}.
Among other things, 
in such a system one can expect a multitude of correlated fermionic 
phases as well as superfluidity. 
In a two-component fermionic gas the atom numbers
of different components can be independently controlled and such
strongly interacting polarized fermion gases
have been recently 
studied experimentally~\cite{Partridge2006a,Zwierlein2006a,Shin2006a}. Such studies have revealed,
for example, intriguing phase separation properties and, depending on
parameters, appearance of trap physics beyond the 
local-density approximation~\cite{Partridge2006c}.

In this article we address the issue of a {\it strongly} polarized two-component
Fermi gas in an optical lattice. Polarized Fermi gases on the lowest band
have been studied previously, see for example Ref.~\cite{Koponen2007a}, but here
we wish to investigate 
the relevant theory as well as phases associated with it, 
when the majority component fills the lowest band and also populates 
excited bands. 
To this end, we 
derive the 
%EL repeated word
%relevant 
Hamiltonian for this system and 
apply it to discuss the possible phases both for repulsive as well as
attractive interactions. 

Multiband physics has a long history in the condensed
matter theory and superconductors with 
several bands have been addressed within the BCS as well as
Ginzburg-Landau formalism~\cite{Suhl1959a,Tilley1964a}. 
However, a system of polarized ultracold
atoms in optical lattices is different in many important
respects. First,  in the theory of transition metals,
for example, the multiband description is motivated by the fact that
the Fermi surface might pass through
two (or more) bands. Under such circumstances there are processes which
transfer electrons between bands as well as interactions inside the bands.
In the system that we consider the bands are well separated
and there are no processes that transfer atoms between bands. Here the interaction
between the bands is proportional to the product of the densities in different bands
and does not involve transfer of atoms between bands~\cite{Suhl1959a}. Second, in the present system
only interband interactions are relevant. This
makes it much easier to study interband effects without
being disturbed by large inband effects. Third, the anisotropic nature
of the $p$-band tunneling as well as  the multi-flavor character
of the $p$-band fermions gives rise to qualitatively novel possibilities.
Fourth, the Hamiltonian
for the system we discuss can be derived from the microscopic theory in a controlled
way.

Higher band physics with Bose-Einstein condensates was experimentally studied
by M\"{u}ller {\it et al.}~\cite{Mueller2007a} and
interaction induced transitions to higher bands were observed
by K\"{ohl} et al.~\cite{Kohl2005a} while Diener and 
%EL
T.-L.~Ho~\cite{Diener2006a}
tackled this  problem theoretically. In this paper, however, the interaction
strengths are much smaller than the bandgaps and such interaction induced
higher band effects do not play an important role.
Other
interesting possibilities also exist. For example, the higher band atoms
might have stronger nearest neighbor interactions which
might make super solid phases appear in some parts of the phase diagram~\cite{Scarola2005a}.
On the other hand when the lattice geometry is varied, unconventionally
ordered quantum phases are predicted to be possible~\cite{CWu2006a,CWu2007a}.
Some aspects of the higher band physics in one dimension
were also discussed by Kantian {\it et al.}~\cite{Kantian2007a} in the context of 
lattice excitons, by K\"{a}rkk\"{a}inen {\it et al.}~\cite{Karkkainen2007a} 
for repulsive interactions with high filling fractions,
and by A.~F.~Ho~\cite{Ho2006a}, who focused on the equal mixture Fermi gas
with strong interactions and with two atoms per site. 
More recently anti-ferromagnetic properties of $p$-band fermions
at half filling were studied by Wu and Zhai~\cite{Wu2007a}
and $p$-band fermions in a two-dimensional lattice with different lattice
geometries by Zhao and Liu~\cite{Zhao2008a}.

\section{Hamiltonian}
Atoms in a cubic optical lattice experience a potential 
$
V_\sigma({\bf r})=\sum_\alpha V_{\alpha,\sigma}\sin^2 \pi r_\alpha/d,
$
where $\alpha=\{x,y,z\}$,  $V_{\alpha,\sigma}$ are the lattice depths for atoms of type $\sigma$, 
and $d$ is the lattice constant.
Our interest is in a strongly polarized two-component gas where the majority component
fills the lowest band and occupies also part of the first excited bands. We label
the components by $\uparrow$ and $\downarrow$. 
A Hubbard-model Hamiltonian for the fermions is arrived at by expanding the 
field operators ${\hat \psi}_\sigma({\bf r})$ in terms of the localized 
Wannier functions~\cite{Jaksch1998a}. 
%EL this had to be shaped up
The Wannier wavefunctions on the lowest band, 
$w_{0,\sigma}({\bf r})$, are even functions, while the Wannier function 
$w_{\alpha,\uparrow}({\bf r})$ on the $p$-band has a node in the plane 
normal to the coordinate axis $\alpha$, giving rise to three types of state
for the spin-up 
fermions which we label $x$, $y$, and $z$, respectively ~\cite{Isacsson2005a}.
Including only the leading nearest neighbor tunneling terms, we find 
the Hamiltonian for the ideal two-component Fermi gas in momentum space 
\begin{eqnarray}
H_{\rm ideal}&=&\sum_{\sigma,\bf k} \left(\epsilon_{0,{\bf k},\sigma}-\mu_\sigma\right)
\psi_{0,{\bf k},\sigma}^\dagger\psi_{0,{\bf k},\sigma}\\
&+&\sum_{\alpha,\bf k}\left(\epsilon_{\alpha,{\bf k},\uparrow}-\mu_\uparrow\right)
\psi_{\alpha,{\bf k},\uparrow}^\dagger\psi_{\alpha,{\bf k},\uparrow}\nonumber,
\end{eqnarray}
where $\sigma=\{\uparrow,\downarrow\}$.
Here the summation is over the first Brillouin zone, $\psi_{0,{\bf k},\sigma}^\dagger$ creates an atom in the lowest band, 
$\mu_\sigma$  are the chemical potentials~\cite{comment2007a}, and  $\epsilon_{0,{\bf k},\sigma}$ and $\epsilon_{\alpha,{\bf k},\sigma}$
are the dispersions on the lowest and excited $p$-bands respectively.
Furthermore, $\psi_{\alpha,{\bf k},\uparrow}^\dagger$ creates
a spin-up atom with momentum ${\bf k}$.

The lowest band dispersion is given by the usual expression
$\epsilon_{0,{\bf k},\sigma}=\sum_{\beta=\{x,y,z\}} 2J_{0,\sigma,\beta}(1-\cos k_\beta d)$, 
where in principle the tunneling strength
can depend on the spin-state as well as on direction if the lattice depth is different in different directions.
The excited band dispersions are more complex. First of all, the excited bands are separated from the lowest
band by  energy gaps $\Delta_{\alpha,BG}$. Second, since the Wannier functions on the excited band
have a node and are antisymmetric along some axis, the tunneling strength for moving an atom
in the direction orthogonal to the nodal plane is different from moving it in the direction along the nodal plane.
The dispersions are then 
$
\epsilon_{\alpha,{\bf k},\uparrow}=\Delta_{\alpha,BG}
+\sum_{\beta=\{x,y,z\}} 2J_{\alpha \beta}\left(1-\cos k_\beta d\right), 
$
where $J_{\alpha \beta}$ is the tunneling strength in the direction $\beta$
for an atom which has a localized wavefunction
$w_\alpha({\bf r})$.

In ultracold gases the dominant interaction between 
unlike fermions is typically the $s$-wave interaction
$
g\int d{\bf r} n_\uparrow({\bf r})n_\downarrow({\bf r}),
$
where the coupling $g$ can be  expressed in terms of the $s$-wave scattering length $a$ and atomic mass
$m$ as $g=4\pi\hbar^2 a/m$. This interaction term can again be reduced into the lattice by expanding the field
operators and keeping only the leading on-site interaction terms. This procedure gives us a contribution
in coordinate space
\begin{eqnarray}
H_I&=&U_0\sum_{{\bf i}=\left( i_x,i_y,i_z\right)} \psi_{0,{\bf i},\uparrow}^\dagger \psi_{0,{\bf i},\downarrow}^\dagger 
\psi_{0,{\bf i},\downarrow} \psi_{0,{\bf i},\uparrow}+\\
&&\sum_{{\bf i}=\left( i_x,i_y,i_z\right)} \sum_{\alpha=\{x,y,z\}}U_{1,\alpha}\psi_{\alpha,{\bf i},\uparrow}^\dagger \psi_{0,{\bf i},\downarrow}^\dagger 
\psi_{0,{\bf i},\downarrow} \psi_{\alpha,{\bf i},\uparrow}\nonumber
\end{eqnarray}
to the Hamiltonian. The coupling strength $U_0$ between atoms on the lowest band
is related to the scattering length through
$
U_0=g\int d{\bf r} |w_{0,\downarrow}({\bf r})|^2|w_{0,\uparrow}({\bf r})|^2
$
while the interband couplings between the $\downarrow$-atoms on the lowest band
and the  $\uparrow$-atoms on the excited band are given by
$
U_{1,\alpha}=g\int d{\bf r} |w_{0,\downarrow}({\bf r})|^2|w_{\alpha,\uparrow}({\bf r})|^2.
$

In principle, the localized Wannier functions could be calculated numerically from the three-dimensional
band structure, but this is 
unnecessary for our purposes. One gets reasonable analytical estimates for all the parameters
of the theory by approximating the Wannier functions with harmonic oscillator states localized at
each lattice site. %which is approximated as a three-dimensional harmonic oscillator. 
General features
are not sensitive to precise numerical values of the parameters and 
%EL
the 
harmonic approximation
gives us a handle on how various parameters vary relative to one another 
as 
the lattice depth 
is changed. The analytical formulas
for the parameters are long and not very informative and are, for that reason, omitted here.
It is however necessary to discuss some general features. 

Firstly, for typical parameters
the bandgaps are much higher than other energy scales of the problem. Second, in the harmonic 
approximation the interband coupling is simply $U_{1,\alpha}=U_0/2$ and is independent of 
which excited band is involved, {\it even} for an-isotropic lattices. Third, on the excited bands
the diagonal tunneling strengths $J_{\alpha\alpha}$ have an opposite sign to the off-diagonal 
strengths (as well as tunneling strengths on the lowest band) and also their magnitude is
much larger than those for the off-diagonal hopping strengths. 
This is a simple consequence of the Wannier function being an odd function 
of the coordinate normal to the nodal plane, as well as having a wider 
extension along that direction. This is also implied by the familar bandstructures
in a one-dimensional system where ther lowest band has a minimum at $k=0$ while the
first excited band has a minimum at the edge of the Brillouin zone.

Because of the strong anisotropy of the excited band tunneling strengths, the structure 
of the ideal gas Fermi surface
is quite unlike that in the lowest band. In the lowest band,
the Fermi surface is roughly spherical for low filling fractions, but on the excited bands
Fermi surfaces are more sheet-like since atoms first fill the states along the directions perpendicular
to the direction of large tunneling strength.
The Hamiltonian we derived is very rich and
the possible phases that can occur in different parameter regimes
are numerous. We now proceed to discuss a few applications. First, 
we discuss anti-ferromagnetic phases with repulsive interactions and then we proceed
to discuss the paired phases with attractive interactions.

\section{Repulsive interactions}

The order parameter for an antiferromagnetic state with an ordering vector 
$\QQ$ is defined as
$
%\label{afop}
A_{\alpha,\QQ,\sigma} = V^{-1}\sum_{\kk} 
\la \psi^{\dagger}_{\alpha,\kk+\QQ,\sigma}
\psi_{\alpha,\kk,\sigma}\ra,
$
where $V$ is the dimensionless volume, i.e., the total number of sites in 
the system.
In principle, two types of antiferromagnetic state are conceivable in the 
repulsive case: either one with checkerboard symmetry, so that 
$\QQ=(\pi,\pi,\pi)$ in Cartesian coordinates, or a striped phase where 
symmetry is broken, $\QQ=(\pi,0,0)$. 
However, as we shall see, the 
striped phase is found to be energetically unfavorable. 
Following 
Ref.~\cite{Mahan}, the interaction part of the 
Hamiltonian is written in a mean-field 
approximation as 
\bea
H_{\rm I}' &=& {\sum_{\alpha\kk}}' U_{1,\alpha} n_{0\downarrow} 
\left(\psi^{\dagger}_{\alpha\kk\uparrow} \psi_{\alpha\kk\uparrow} 
+ 
\psi^{\dagger}_{\alpha,\kk+\QQ,\uparrow} \psi_{\alpha,\kk+\QQ,\uparrow} 
\right)
\nonumber\\
&+&  
U_{1,\alpha} n_{\alpha\uparrow} \left(
\psi^{\dagger}_{0,\kk\downarrow} \psi_{0,\kk\downarrow}
+
\psi^{\dagger}_{0,\kk+\QQ,\downarrow} \psi_{0,\kk+\QQ,\downarrow}
\right)
\nonumber\\
&+& 
U_{1,\alpha}A_{0,\QQ,\downarrow} \left(\psi^{\dagger}_{\alpha,\kk+\QQ,\uparrow}\psi_{\alpha\kk\uparrow} + {\rm h.c.}\right) 
\nonumber\\
&+& 
U_{1,\alpha}A_{\alpha,\QQ,\uparrow} \left(\psi^{\dagger}_{0,\kk+\QQ,\downarrow}\psi_{0,\kk,\downarrow} + {\rm h.c.}\right) 
\nonumber\\
&-& V \sum_{\alpha}U_{1,\alpha} \left( n_{0\downarrow}n_{\alpha\uparrow} 
+ A_{0,\QQ\downarrow}A_{\alpha\QQ\uparrow} 
\right),
\ena
where the primed sum extends only over the reduced Brillouin zone (RBZ), 
defined such that the points $\{\kk, \kk+\QQ | \kk \in {\rm RBZ}\}$ 
make up the entire first Brillouin zone. The quantities $n_{0\downarrow}$ 
and $n_{\alpha\uparrow}$ are the densities of the components occupying 
the respective orbitals. A Bogoliubov transformation 
diagonalizes the Hamiltonian, whereafter the free energy 
$
\Omega(A_{0,\QQ,\downarrow},A_{x,\QQ,\uparrow},A_{y,\QQ,\uparrow},A_{z,\QQ,\uparrow})=-k_BT\log\{{\rm Tr}\,\exp[-\beta (H_{\rm ideal}+H_{\rm I}')]\}
$
can be computed. Here, $\beta=1/(k_BT)$, and $T$ is the temperature. 
For definiteness, we assume that both components experience
the same lattice potential and that the lattice potential has the same
depth in all directions. 
Parameters are calculated for $^{40}{\rm K}$ atoms in a lattice
of depth $V_0=8\, E_R$, where $E_R$ is the recoil energy of the 
atoms when they absorb a photon (of wavelength $826\, {\rm nm}$).
In that case, we obtain $J_0=0.012E_R$, $J_{\alpha\alpha} = -0.15E_R$, and
$J_{\alpha\beta}=0.0028E_R$ if $\alpha\neq\beta$, 
which demonstrates our earlier qualitative arguments about the magnitude
of the tunneling terms.  
For $a=174$ Bohr radii, $U_0=0.30E_R$ \cite{Regal2003b}.
We assume that using a magnetic field $a$ can be tuned 
to arbitrary positive and negative values~\cite{Kohl2005a}. %; in the absence of a magnetic field it is negative. 

\begin{figure}
\includegraphics[width=0.90\columnwidth]{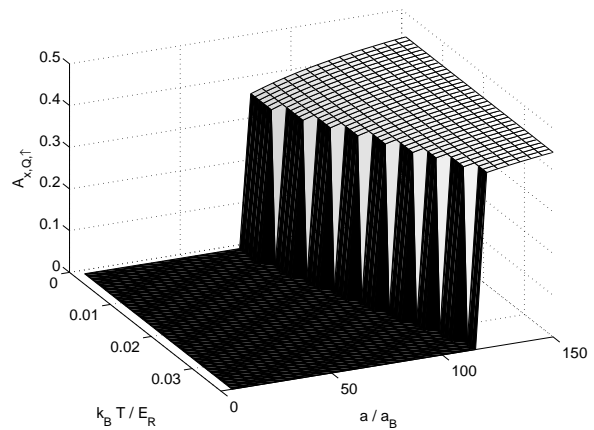} 
\caption{
Magnitude of the antiferromagnetic order parameter $A_{x\QQ\uparrow}$ 
for the majority component 
in the checkerboard state, $\QQ=(\pi,\pi,\pi)$, as a 
function of temperature and scattering length for a gas of $^{40}$K atoms 
in a lattice with strength $V_0=8E_R$, where $E_R$ is the recoil energy. 
The scattering length $a$ is given 
in units of the Bohr radius $a_B$.
}
\label{fig:af}
\end{figure}

As $a$ is increased, the system undergoes a first-order transition from 
the normal state, with $A_{\alpha\QQ\sigma}=0$ for all $\alpha$, to a 
checkerboard state where $A_{\alpha\QQ\sigma}\neq 0$ and 
$A_{x,\QQ,\uparrow}=A_{y,\QQ,\uparrow}=A_{z,\QQ,\uparrow}$. The order 
parameter 
$A_{x\QQ\uparrow}$ is plotted in Fig.\ \ref{fig:af}. 
The magnitude of $A_{0,\QQ,\downarrow}$ displays a similar behavior. 
The chemical 
potentials are in the calculation fixed to 
$\mu_{\downarrow}=6J_{0,\downarrow}$ 
and $\mu_{\uparrow}=\Delta_{x,BG}+2(J_{xy}+J_{xz})$, respectively; 
this is in the noninteracting 
limit close to the value for half-filling. 
The phase diagram is found to 
be very insensitive to the exact values of the chemical potentials; in the 
antiferromagnetic state, the densities are locked to the 
values $n_{\downarrow}=0.5$ and $n_{\uparrow}=1.5$, respectively, so 
that for the 
majority component, 
the mean occupation of the $0$ orbital is 
unity and the combined occupation of the $x$, $y$, and $z$ orbitals is 0.5. 
If the fermions are confined in a magnetic trap, the state of the 
system is to a good approximation given by a local-density approximation 
where the local chemical potential is given by the sum of the 
chemical potential and the negative of the trapping potential. The fact 
that the state of the system is insensitive to the chemical potential 
then 
means that without much fine-tuning, the antiferromagnetic state can be made 
to occupy a large area in 
the trap, with corrections only toward the edges where the density drops 
to zero.

A fully antiferromagnetic state, where the occupations of neighboring 
states in each spin state alternates between 0 and 1, has an order 
parameter of magnitude exactly equal to 0.5. It is seen in Fig.\ \ref{fig:af} 
that the system in the checkerboard state is always close to this 
limit because of the strong repulsive interactions.
The absence of a striped phase in these calculations is not conclusive 
proof that such a state is always thermodynamically unfavorable. On the 
contrary, it is conceivable that striped phases could show up, e.g., 
in anisotropic lattices or in lower dimensions. However, for the parameter 
ranges that we investigated here, the checkerboard state always has the 
lower free energy.

\section{Attractive interactions}
In order to study BCS-type paired states with attractive interactions (i.e. negative $a$),
we introduce auxiliary (pairing) fields
$\Delta_0=U_0 \la \psi_{0,{\bf i},\downarrow}\psi_{0,{\bf i},\uparrow}\ra$ and
$\Delta_\alpha=U_{1,\alpha}\la \psi_{0,i,\downarrow}\psi_{\alpha,i,\uparrow}\ra$
which we use to decouple the interaction terms in the usual way. In this way
we find the interaction term of the mean-field Hamiltonian
\begin{eqnarray}
H_{\rm I}''&=&\sum_{\bf i} \Delta_0 \psi_{0,{\bf i},\uparrow}^\dagger \psi_{0,{\bf i},\downarrow}^\dagger 
+\Delta_0^*\psi_{0,{\bf i},\downarrow} \psi_{0,{\bf i},\uparrow} -|\Delta_0|^2/U_0
\nonumber\\
&+&\sum_{\alpha,\bf i}
\Delta_\alpha\psi_{\alpha,{\bf i},\uparrow}^\dagger \psi_{0,{\bf i},\downarrow}^\dagger 
+\Delta_\alpha^*\psi_{0,{\bf i},\downarrow} \psi_{\alpha,{\bf i},\uparrow} -|\Delta_\alpha|^2/U_{1,\alpha}\nonumber
\end{eqnarray}
The Hamiltonian 
is then
diagonalized with a canonical transformation and 
the grand potential 
%$\Omega(\Delta_0,\Delta_x,\Delta_y,\Delta_z)$
computed and minimized in order to find the state
that is physically realized.

\begin{figure}
\includegraphics[width=0.90\columnwidth]{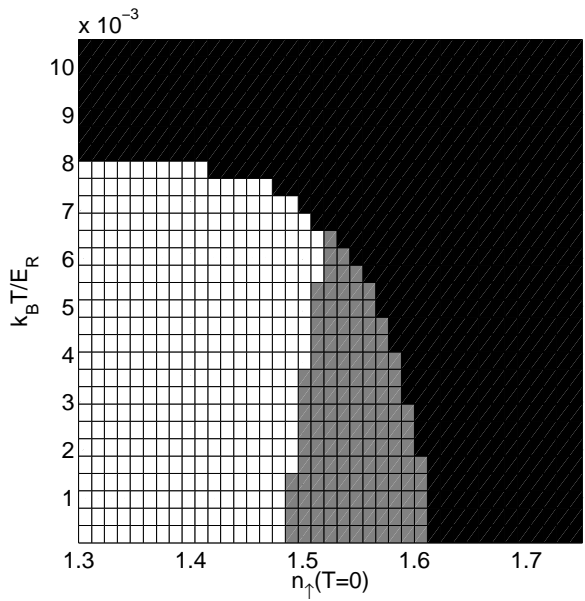} 
\caption[Fig1]{%(color online)
Phase diagram with a fixed scattering length $a=-80a_B$
and
the minority component chemical potential which 
corresponds to half-filling for the ideal system at $T=0$. 
The $x$-axis shows the filling fraction of the 
ideal gas of majority atoms at $T=0$ 
Shading: symmetric state = light, on-axis state = gray, 
normal state=dark.
}
\label{fig:phasediag}
\end{figure}

In Fig.~\ref{fig:phasediag} we show an example phase diagram when the
minority component chemical potential is fixed to 
%EL
the 
value corresponding
to half filling for an ideal Fermi gas at $T=0$ and the majority component chemical potential
and temperature are varied. The lattice parameters were chosen 
the same as before. For the assumed symmetric lattice we can identify three 
different phases: the normal
state where all pairing fields vanish, the on-axis state where only one
$\Delta_\alpha\neq 0$, and the symmetric state where all pairing fields are equal,
i.e., $\Delta_x=\Delta_y=\Delta_z$. The last one of these states dominates at low
temperatures and for lower majority component filling factors. The on-axis state
can be favorable at low temperatures and somewhat higher filling factors, while
the normal state is favorable elsewhere. It should be 
noted, 
that whether
or not the on-axis state appears depends also on the coupling strength. If the
coupling is increased, the symmetric state with equal pairing fields occurs
on a larger part of the phase diagram.

>From our computations (not shown in the figures) 
we observe 
the presence of a critical coupling strength 
before pairing can take place. 
The reason for the critical coupling strength has to do with a different 
structure of the Fermi surfaces of the majority and minority atoms. For identical atoms at the lowest
band 
%EL
the 
Fermi surfaces can be perfectly matched for zero polarization, but this is no longer true
when 
%EL
one of the components 
occupies states on the excited $p$-bands. In this case sufficiently strong
coupling is required to counteract the effect of the mismatched Fermi surfaces. 
Just above the critical coupling strength the on-axis state has a slightly
lower energy than the symmetric state while for stronger couplings the latter
state is favored. 

%EL
\section{Conclusions}
Optical lattice experiments usually include a parabolic trapping potential. When
one starts to increase the number of majority atoms while keeping the number of
minority atoms fixed, the higher bands will not be occupied straight away. At first,
the cloud of majority atoms spreads out in the harmonic trap and their filling factor
in the optical lattice remains less than one. However, roughly at distances larger than $R$ the 
energy $m\omega_T^2R^2/2$ due to the trapping potential with frequency $\omega_T$ becomes
larger than the bandgap. When this happens, it is favorable for the atoms to start
filling the excited band(s) in the center of the trap. This implies that
in a trapping environment the phases discussed in this paper can occur
in the center of the atomic cloud and that this center will be surrounded by
a cloud of majority atoms occupying the lowest band. 
The presence of the pairing gaps and anti-ferromagnetic ordering
would be 
observable, for example, through noise-correlation 
experiments~\cite{Altman2004a,Foelling2005a,Greiner2005b}.
In the noise correlation experiments the structure of the correlation peaks will depend
on the symmetry of initial Wannier functions of the atoms prior to free expansion and can 
therefore be used to distinguish various phases.
Pairing gaps could also be observed through radio-frequency 
sprectroscopy~\cite{Chin2004a,Kinnunen2004b,Shin2007a}.

In this article we derived a theory for the two-component polarized fermions in an optical lattice
when also the lowest excited $p$-bands are occupied. Based on this theory we studied
anti-ferromagnetic phases as well as mean-field BCS-type theory with several possible
order parameters. We 
outlined
the expected phase boundaries for anti-ferromagnetic phases, whose properties 
need to be studied in greater detail in the future. 
In the attractive case we assumed BCS order parameters which do not break translational symmetry, thus postponing the investigation of states which do break the translational symmetry
in a lattice~\cite{Sheehy2007a,Koponen2007a}.
Interesting physics is also expected when the minority component starts to populate the excited
bands. In this case one can expect competition between inter-band and intra-band effects.

{\it Acknowledgments} This work was supported by Academy of Finland (project number 207083), and 
by the Swedish Research Council.

\bibliographystyle{apsrev}
%\bibliography{bibli}

\begin{thebibliography}{31}
\bibitem[{com()}]{comment2007a}
\bibinfo{note}{Note that in using a single chemical potential for the majority
  atoms we fix the total majority component atom number. One could also, in
  priciple, conserve the atom number of each component of the majority atoms
  separately, in which case one should use a separate chemical potential for
  each component.}
\end{thebibliography}

\end{document}